\documentclass[]{jkas} 

\def\year{2024} 
\def\volume{?} 
\def\issue{?} 
\def\beginpage{1} 
\def\received{---} 
\def\accepted{---} 
\def\published{---} 
\date{Received \received; Accepted \accepted; Published \published}

\title{%
A New Collision Avoidance Fiber Assignment Algorithm for Robotic Fiber Positioners in Multi-Object Spectroscopy 
}

\author[1]{Minseong Kwon}{0009-0007-9451-5337}
\author[1,2,3,4]{Ho Seong Hwang}{0000-0003-3428-7612}
\author[3]{Jong Chul Lee}{}
\author[3]{Jae-Woo Kim}{}
\author[1]{Hyeonguk Bahk}{0009-0002-9878-1126}
\author[5]{Young-Man Choi}{}
\author[3]{Moo-Young Chun}{}
\author[3]{Sang-Hyun Chun}{} 
\author[6]{Haeun Chung}{}
\author[3,7]{Sungwook E. Hong}{}
\author[8,3]{Minhee Hyun}{0000-0003-4738-4251}
\author[9,10]{Donghui Jeong}{}
\author[3]{Kang-Min Kim}{}
\author[12]{Dachan Kim}{} 
\author[1,11]{Dongkok Kim}{0000-0003-4127-6110}
\author[3,7]{Yunjong Kim}{}
\author[3]{Jongwan Ko}{}
\author[3]{Ho-Gyu Lee}{0000-0002-3808-7143}
\author[3]{Yongseok Lee}{}
\author[5]{Hyunho Lim}{}
\author[3]{Heeyoung Oh}{0000-0002-0418-5335}
\author[9]{Changbom Park}{0000-0001-9521-6397}
\author[12]{Hyunmi Song}{0000-0002-4362-4070}
\author[7,3]{Mingyeong Yang}{}
\author[13]{Yongmin Yoon}{}

\affil[1]{Astronomy Program, Department of Physics and Astronomy, Seoul National University, 1 Gwanak-ro, Gwanak-gu, Seoul 08826, Republic of Korea}
\affil[2]{SNU Astronomy Research Center, Seoul National University, 1 Gwanak-ro, Gwanak-gu, Seoul 08826, Republic of Korea}
\affil[3]{Korea Astronomy and Space Science Institute, 776 Daedeok-daero, Yuseong-gu, Daejeon 34055, Republic of Korea}
\affil[4]{Australian Astronomical Optics - Macquarie University, 105 Delhi Road, North Ryde, NSW 2113, Australia}
\affil[5]{Department of Mechanical Engineering, Ajou University, 206 Worldcup-ro, Yeongtong-gu, Suwon-si, Gyeonggi-do, 16499, Republic of Korea}
\affil[6]{University of Arizona, Steward Observatory, 933 N. Cherry Avenue, Tucson, AZ 85721, USA}
\affil[7]{Astronomy Campus, University of Science and Technology, 776 Daedeok-daero, Yuseong-gu, Daejeon 34055, Republic of Korea}
\affil[8]{Kavli Institute for Particle Astrophysics and Cosmology, Stanford University, Menlo Park, CA 94305, USA}
\affil[9]{School of Physics, Korea Institute for Advanced Study (KIAS), 85 Hoegiro, Dongdaemun-gu, Seoul 02455, Republic of Korea}
\affil[10]{Department of Astronomy and Astrophysics, and Institute for Gravitation and the Cosmos, The Pennsylvania State University, University Park, PA 16802, USA}
\affil[11]{Institute for Data Innovation in Science, Seoul National University, Seoul 08826, Korea}
\affil[12]{Department of Astronomy and Space Science, Chungnam National University, Daejeon 34134, Republic of Korea}
\affil[13]{Department of Astronomy and Atmospheric Sciences, Kyungpook National University, Daegu 41566, Republic of Korea}

\def\corrauthor{%
Ho Seong Hwang, \email{hhwang@astro.snu.ac.kr}
}

\def\runningauthor{%
Kwon et al.
}

\def\runningtitle{%
A New Fiber Assignment Algorithm
}

\def\keywords{%
software: public release --- surveys --- instrumentation: spectrographs --- methods: observational --- techniques: spectroscopic --- galaxies: general
}

\def\abstracttext{%
We present a new fiber assignment algorithm for a robotic fiber positioner system in multi-object spectroscopy.
Modern fiber positioner systems typically have overlapping patrol regions, resulting in the number of observable targets being highly dependent on the fiber assignment scheme.
To maximize observable targets without fiber collisions, the algorithm proceeds in three steps. 
First, it assigns the maximum number of targets for a given field of view without considering any collisions between fiber positioners.
Then, the fibers in collision are grouped, and the algorithm finds the optimal solution resolving the collision problem within each group. 
We compare the results from this new algorithm with those from a simple algorithm that assigns targets in descending order of their rank by considering collisions.
As a result, we could increase the overall completeness of target assignments by $10\%$ with this new algorithm in comparison with the case using the simple algorithm in a field with 150 fibers.
Our new algorithm is designed for the All-sky SPECtroscopic survey of nearby galaxies (A-SPEC) based on the K-SPEC spectrograph system, but can also be applied to similar fiber-based systems with heavily overlapping fiber positioners.
}

\begin{document}
\jkashead 


\section{Introduction\label{sec:intro}}
\subsection{Spectroscopic Surveys with Multi-object spectrographs}

Galaxy redshift surveys have played a crucial role in advancing our understanding of galaxies and large-scale structures in the universe (e.g. \citealt{1991ARA&A..29..499G, 2004LRR.....7....8L}). Following the pioneering survey by \citet{1983ApJS...52...89H}, many other surveys have provided comprehensive measurements of galaxy redshifts, allowing us to study galaxy evolution \citep[e.g.][]{2009ApJ...700..791H,2014ApJ...791..130Z} and large-scale structures in the universe \citep[e.g.][]{1986ApJ...302L...1D,1989Sci...246..897G,2005ApJ...624..463G}. 
These redshift surveys have been revolutionized by multi-object spectroscopy (i.e. MOS), which enables the simultaneous acquisition of spectra for hundreds to thousands of objects. MOS systems achieve high multiplexing and flexibility through techniques such as multi-slit masks that use custom-designed plates or masks for each field \citep{1990MNRAS.244..408C,2004PASP..116..425H}, fiber plug-plates \citep{2013AJ....146...32S}, as well as robotic fiber positioners that allow rapid reconfiguration, thereby making them well suited for large surveys \citep{1996ApJ...470..172S,2005PASP..117.1411F,2021AJ....161...92S}.
Some notable wide-field surveys using MOS include the Las Campanas Redshift Survey \cite[LCRS,][]{1996ApJ...470..172S}, 2-degree Field Galaxy Redshift Survey \cite[2dFGRS,][]{2001MNRAS.328.1039C}, 6-degree Field Galaxy Survey \cite[6dFGS,][]{2004MNRAS.355..747J}, and the Sloan Digital Sky Survey \cite[SDSS,][]{2000AJ....120.1579Y}.
Each of these surveys provides redshifts of more than 100,000 galaxies, enabling statistical analysis of structure formation and cosmology \citep[e.g.][]{2007ApJ...658..898P,2023ApJ...953...98D}.

Recent initiatives, such as the Dark Energy Spectroscopic Instrument \cite[DESI,][]{2016arXiv161100036D} and the James Webb Space Telescope (JWST, \citealt{2023PASP..135f8001G}, see \citealt{2022A&A...661A..81F} for its MOS capability), continue to provide important data from the early to the late universe. However, these surveys are still far from being complete in terms of areal coverage and apparent magnitudes of galaxies, necessitating further observations. 

\subsection{A-SPEC and K-SPEC}
\begin{figure*}[t]
    \centering
    \includegraphics[width=1\textwidth]{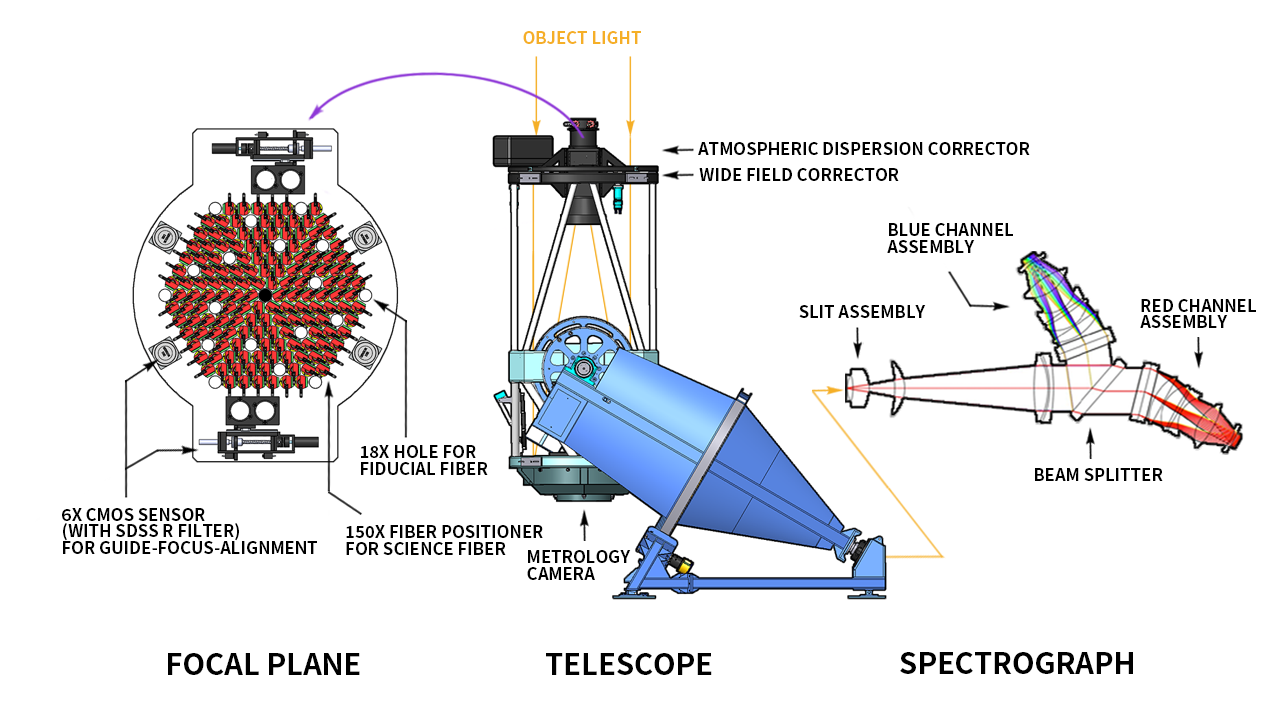}
    \caption{A schematic view of K-SPEC system}
    \label{fig: kspec}
\end{figure*}
The measurement of the Hubble constant, which represents the expansion rate of the universe at the current epoch, has gradually evolved since Edwin Hubble's first estimate in 1929 \citep[e.g.][]{2021CQGra..38o3001D}. There are two types of measurements: an indirect method that finds the best-fit model parameters through the analysis of the cosmic microwave background (CMB) from the distant universe \citep[e.g.][]{2021A&A...652C...4P}, and a direct method that derives from galaxy data, using radial velocity and distance measurements in the local universe \citep[e.g.][]{2022ApJ...934L...7R}. Interestingly, recent measurements from these two independent methods reveal a statistically significant difference \cite[so-called Hubble tension; ][]{2022NewAR..9501659P}. Various explanations have been proposed to address this tension \citep{2020ApJ...905..104K,2020PhRvD.101d3533K}. One possible solution for this tension is to accept the discrepancy between the two measurements and seek an explanation for it; e.g. local measurement of the Hubble constant could be higher than the mean value of the universe from the CMB if we reside in an underdense region \citep{2014MNRAS.440..119L,PhysRevD.93.023529}. To examine this idea and related models, it is crucial to accurately measure the matter distribution around us at $z \lesssim 0.1$, where the density is expected to converge to the mean value \citep{2020A&A...633A..19B}. 
This requires all-sky survey data covering a large volume of the local universe. The 2MASS Redshift Survey \cite[2MRS,][]{2012ApJS..199...26H} provides a complete sample of galaxies with $K_S \leq 11.75$, but its magnitude limit is not deep enough for this purpose. We therefore need an all-sky spectroscopic survey with a fainter magnitude limit. This will allow us to study the evolution of dark energy by combining data across different redshifts \citep[e.g.][]{2023ApJ...953...98D} and to investigate the physical properties of nearby galaxies by integrating existing multi-wavelength all-sky survey data (e.g. SPHEREx: \citealt{2016arXiv160607039D}, WISE: \citealt{2010AJ....140.1868W}). In this regard, we are planning a new All-sky SPECtroscopic survey of nearby galaxies (A-SPEC) to construct a highly complete sample of galaxies with measured redshifts down to a magnitude of $K_S \leq 13.75$. This limiting magnitude is 2 magnitudes fainter than that of 2MRS, resulting in a sample nearly 20 times larger. To do that, we select $\sim 780,000$ nearby galaxies with $K_S \leq 13.75$ from the 2MASS Extended Source Catalog \citep{2000AJ....119.2498J}. Among them, we focus on $\sim 450,000$ galaxies without redshift information in the literature, which constitute our primary targets.
To conduct an all-sky survey with high target density, we develop a new multi-object spectroscopic instrument with robotic fiber positioners called K-SPEC (see Figure~\ref{fig: kspec}). This instrument requires a fiber assignment algorithm which can effectively manage fiber configurations, especially in regions with high target density.

\subsection{Fiber Assignment Algorithm}
In many recent robotic fiber positioner systems for multi-object spectroscopy, the patrol regions of neighboring fiber positioners are designed to overlap significantly so that many fibers can be assigned even in regions with high target density \citep{2018MNRAS.481.3070H, 2021AJ....161...92S}. Therefore, targets should be carefully assigned to fully utilize the mechanical capability of the instrument. 
Sophisticated algorithms have been developed to optimize target assignment. 
For example, the remote swap algorithm recursively tests the exchange of a target assigned to a nearby fiber with an unassigned one, making it effective for radially arranged robotic fiber positioners such as Hectospec \citep{1998SPIE.3355..324R}. Stochastic approaches, such as random subpriority combined with global dithering, allow unbiased estimation of galaxy clustering properties \citep{2019MNRAS.484.1285S}. Another example is cost function optimization, which is computationally efficient and can improve the convergence rate using a specifically designed cost function \citep{2020A&C....3000364M}. However, none of these methods can be directly applied to our K-SPEC system because of its distinct configuration and survey requirements. 
We therefore develop a new open-source fiber assignment algorithm specifically designed for fiber-fed multi-object spectrographs with heavily overlapping workspaces, which addresses fiber collisions by localizing fiber groups and deterministically computing the optimal solution.\footnote{\url{https://github.com/MinseongAstro/KSPEC_FiberAssign}}

This paper is organized as follows. In Section~\ref{sec : kspec}, we introduce the specification of K-SPEC, the multi-object spectrograph system designed for A-SPEC, and present the formulation of fiber collisions for the algorithm.
In Section~\ref{sec : method}, we outline the priorities considered in the algorithm and describe each step in detail.
Finally, in Section~\ref{sec : results} we present example assignments and the results of performance tests for the algorithm.

\section{K-SPEC focal plane system layout}\label{sec : kspec}

\subsection{Robotic fiber positioner}

\begin{table*}[t!]
\caption{Survey characteristics and instrument specifications of the K-SPEC fiber positioner}
\label{tab:jkastable1}
\centering

\renewcommand{\arraystretch}{1.1}
\makebox[\textwidth][c]{%
\begin{tabularx}{1.05\textwidth}{p{3.8cm}p{5.1cm}p{3.2cm}p{5.2cm}}
\hline
\hline
\multicolumn{2}{l}{\textbf{Survey characteristics}} 
& \multicolumn{2}{l}{\textbf{Instrument specifications}} \\
\hline
Telescope
& KMTNet SSO $1.6~\mathrm{m}$
& Fiber angular diameter
& $75~\mu m~\left(3~\mathrm{arcsec}\right)$\\

Field of View per tile
& $5.3~\text{deg}^2$
& Alpha arm length 
& $5.2 \,\mathrm{mm}$ \\

Number of fiber positioners
& $150$ ($130$ targets, $8$ standards, $12$ sky)
& Beta arm length 
& $11.6\,\mathrm{mm}$ \\

Depth
& $K_S < 13.75$ (Vega; $\text{S/N} >5$)
& Patrol radius range
& $6.4-16.8~\mathrm{mm}~\left(2.13-5.60 ~\mathrm{arcmin}\right)$ \\

Wavelength coverage 
& $3700-8700\,$\AA$~\left(R=1800-2100\right)$
& Base position distance 
& $16.8~\mathrm{mm}$ \\

Exposure time
& 15 min on source + 3 min overhead
& Collision buffer, $\sigma_{cb}$ 
& $3.5~\mathrm{mm}$ \\
\hline
\hline
\end{tabularx}
}
\end{table*}


\begin{figure}[t]
    \centering
    \includegraphics[width=0.5\textwidth]{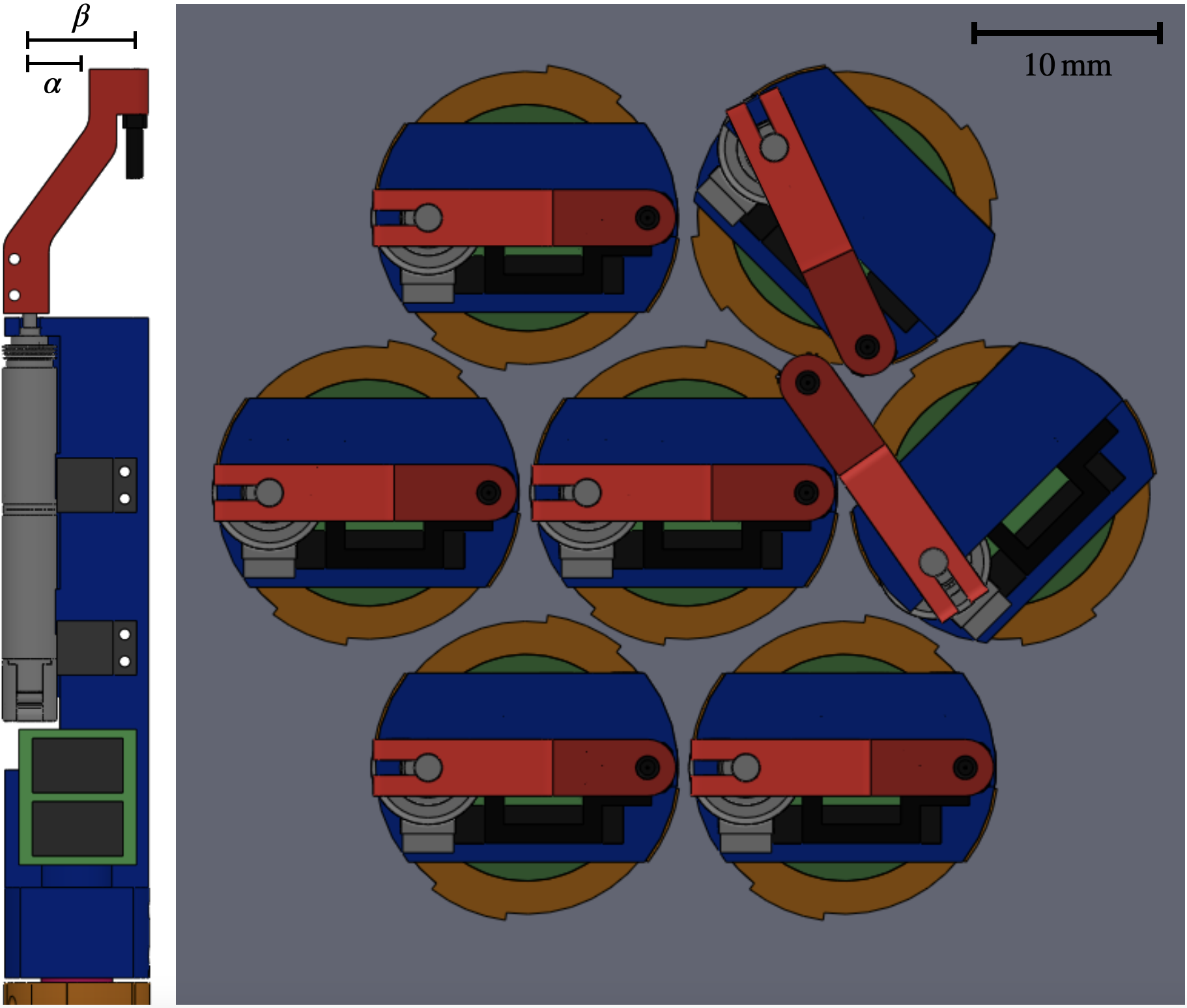}
    \caption{K-SPEC robotic fiber positioners. $\it left$: Side view of a single-unit fiber positioner, which comprises a beta (upper) arm and an alpha (lower) arm that rotate independently. $\it right$: Top view of a collection of seven fibers arranged in a hexagonal array.}
    \label{fig: fiber positioner}
\end{figure}

\begin{figure}[t]
    \centering
    \includegraphics[width=0.5\textwidth]{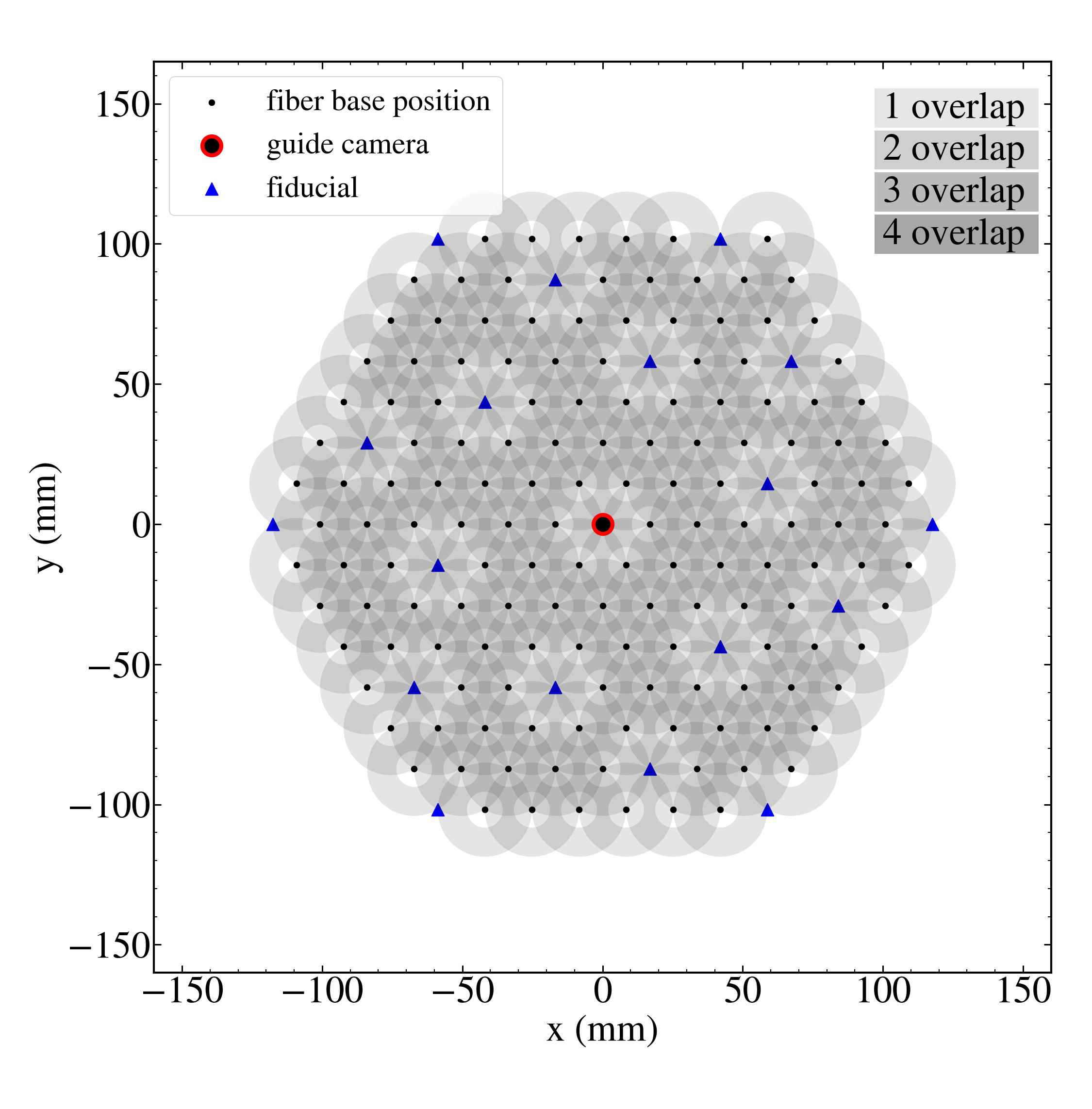}
    \caption{Configuration of K-SPEC fiber positioners. The black dots represent base positions of the robotic fiber positioners. The gray-shaded regions denote the patrol areas of fiber positioners, with varying opacities indicating different levels of overlap. The blue triangles and the red circle represent the fiducials and guide camera, respectively.
    }
    \label{fig: layout}
\end{figure}
Figure~\ref{fig: fiber positioner} shows the side view of a fiber positioner (left panel), and the top view of a pack of seven positioners (right panel) in K-SPEC \citep{2025Lim}. The optical fiber is mounted at the end of each beta arm, and the positioning of the fibers is achieved through the independent rotation of the two arms (i.e. alpha and beta) around their respective axes, similar to the systems in SDSS-V \citep{2021AJ....161...92S} and DESI \citep{2023AJ....165....9S}. The distance between the base positions of the robotic fiber positioners is larger than twice the length of the alpha arm plus the beta arm motor radius, thereby preventing collisions between the alpha arms. Collisions between beta and alpha arms are avoided because their distances from the plate are different (i.e. they are located on different planes). Therefore, we need to consider the collision problem among the beta arms only in the fiber assignment algorithm. Figure~\ref{fig: layout} shows the entire view of the focal plane highlighting the arrangement and patrol region of each robotic fiber. A total of 150 fibers are available for object and sky observations, after excluding 18 fiducials\footnote{Here, we ignore 12 additional fiducials outside the region that the fibers can reach.} (blue triangles) and one central position for the guiding camera (red circle); here, fiducials are light sources distributed across the focal plate field, serving as reference positions for target fiber positioning. Key specifications of the fiber positioners and their configuration in the focal plane are summarized in Table~\ref{tab:jkastable1}.

For K-SPEC, we adopt $l_\alpha= \ 5.2 \ \rm{mm}$ and $l_\beta=\ 11.6 \rm \ mm$ for the lengths of the alpha and beta arms, respectively. Because the beta arm is longer than the alpha arm, the patrol region of the robotic fiber positioner forms an annular area with inner and outer radii of $l_\beta - l_\alpha$ and $ l_\beta+ l_\alpha$, respectively. Moreover, the distance between the two fiber base points is $16.8 \rm \ mm$, which is equal to the sum of the alpha and beta arm lengths. This configuration covers the entire focal plane without gaps, but results in significant overlap among the patrol regions of the fiber positioners; typically, three to four positioners cover each location. Therefore, the number of observable targets in a single configuration significantly depends on the assignment strategy (i.e. which fiber should be used for which galaxy), necessitating a fiber assignment algorithm for optimal target assignment.
In this study, we consider only the right-armed configuration to avoid degeneracy. In this set-up, the alpha arm can rotate from 0 to 360 degrees. However, the beta arm is restricted to counterclockwise rotation, ranging from 0 to 180 degrees from the extended end of the alpha arm.

\subsection{Collision formulation}
\begin{figure}[t]
    \centering
    \includegraphics[width=0.45\textwidth]{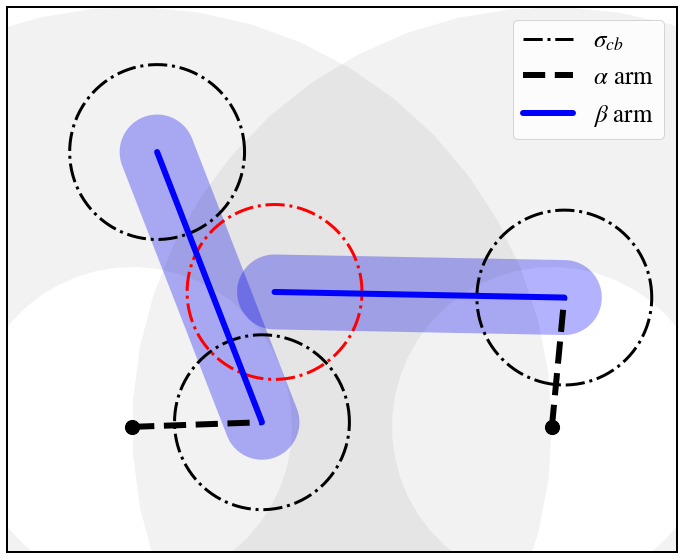}
    \caption{A schematic diagram that illustrates the fiber collision. The dashed and solid lines represent the alpha and beta arms, respectively. The blue shades describe the width of beta arms. The red dot-dashed line indicates the collision buffer $\left(\sigma_{cb}\right)$ of a single fiber positioner, which collides with another fiber positioner.}
    \label{fig: collision}
\end{figure}

\begin{figure*}[t]
    \centering
    \includegraphics[width=1\textwidth]{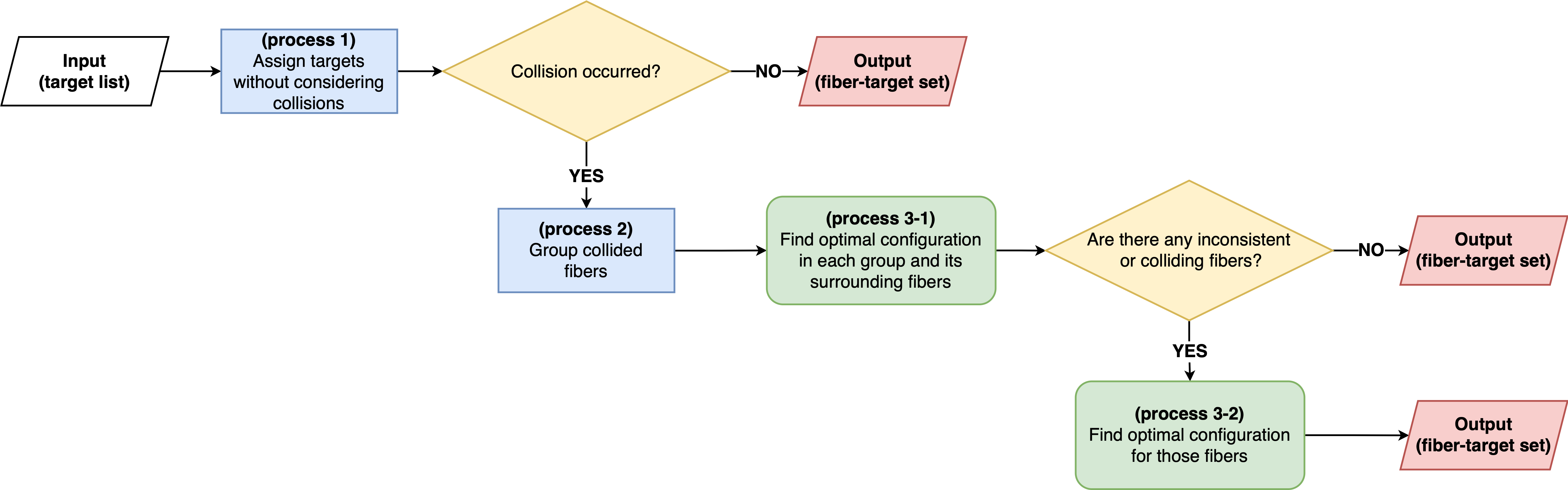}
    \caption{A flowchart of the algorithm.}
    \label{fig: flow}
\end{figure*}

To simplify the collision problem, it is convenient to treat beta arms as line segments in a single plane. If the two line segments do not intersect, the shortest distance between them is the minimum distance from either endpoint to the other segment. Therefore, to determine whether a collision occurs between beta arms, it is sufficient to check whether the line segments intersect and whether the distance from each endpoint to the other segment exceeds a certain threshold. The collision buffer $\left(\sigma_{cb}\right)$ is defined as the lower limit of the shortest distance between the two line segments.

The right panel of Figure~\ref{fig: fiber positioner} shows that the two ends of the beta arm for the K-SPEC positioners have different shapes: one circular and one rectangular. The rectangular ends do not collide with each other, as each is positioned on a different plane. For K-SPEC, we set $\sigma_{cb}=3.5 \ \rm{mm}$ at both ends by considering the width of the beta arm and an additional tolerance; the beta arm has a width of $3\rm \ mm$ and a radius of $1.5 \rm \ mm$ at the circular end.
Figure~\ref{fig: collision} shows an example of this formulation, where two robotic fiber positioners are considered to be in collision because one line segment crosses the collision buffer of another positioner.

\section{Method}\label{sec : method}
The main objective of the algorithm is to assign as many primary targets as possible without fiber collisions.
Here, we explain how to solve the collision problem while maximizing the number of assigned fibers. The algorithm prioritizes the following conditions in order:

\begin{enumerate}
    \item Each fiber positioner should not collide with others.
    \item The number of assigned targets should be maximized.
    \item Targets with higher priorities should be assigned as much as possible.
    \item Computing time should be minimized.
    \item The robotic fiber positioners should be uniformly distributed.
\end{enumerate}
For the optimal target assignment, the algorithm is divided into three processes. First, we select as many fiber-target pairs as possible without considering fiber collisions. We then identify fibers in collision and group them. Finally, we resolve these collisions within each group and assign targets to the fibers. The flowchart of this algorithm is provided in Figure~\ref{fig: flow}. We explain the details of the processes in the following sections.

\subsection{Process 1: Target selection without considering collisions}\label{sec:3.1}
In the first step, we disregard any collisions between robotic fiber positioners and select as many fiber-target pairs as possible. As Figure~\ref{fig: layout} illustrates, the patrol regions of the robotic fiber positioners for K-SPEC significantly overlap, making the optimal target assignment inefficient if we consider the assignment methods based solely on targets or fibers. In this process, the algorithm first identifies all possible fiber-target pairs. Next, it selects a fiber-target pair according to specific conditions. Once the selection is made, the pairs including the selected fiber or target are removed from the list of initial pairs. This selection and removal procedure is repeated until all initial pairs have been removed.
The selection conditions are as follows.

\begin{enumerate}
    \item The target with the fewest assignable fibers is assigned first. If multiple targets have the same number of assignable fibers, the one with the higher rank is assigned.
    \item Among the fibers that can be assigned to the target, the fiber with the fewest assignable targets is assigned to the target. 
    \item If multiple fibers can be assigned to a given target, the fiber that can be most uniformly distributed is used.
\end{enumerate}
We find such a fiber $j$ that minimizes the following equation:
\begin{equation}
\label{eq1}
\sum_{i}{\frac{1}{\left|\mathbf{f}_i-\mathbf{f}_{j}\right|^2}},\quad  \mathrm{where} \ \mathbf{f}_i,\mathbf{f}_j \in \mathbb{R}^2.
\end{equation}
Here, $\mathbf{f}$ represents the fiber base position vector, $\mathbf{f}_i$ is a selected fiber, and $\mathbf{f}_{j}$ is a candidate fiber position vector. 
The rank denotes the priority of a target; for K-SPEC, brighter targets with smaller $K_S$ magnitudes are given higher ranks.
This method can be easily applied to any heavily overlapped fiber configuration when the collision is ignored, and is computationally efficient.

\subsection{Process 2: Group collided fiber positioners}
In the second process, we group fibers that collide with others to localize the problem. 
If a fiber belongs to multiple groups, these groups are merged into a single larger group. Additionally, if there are any unused nearby fibers that can reach a target within the selected fiber-target pairs in the group, they are added to the group. The probability of assigning more targets---and consequently, the computational load---increases with the size of the group. We empirically determine the minimum group size to maximize the number of assigned fibers, using the target candidates of K-SPEC.

\subsection{Process 3: Optimal target assignment in each group}
In the third process, we aim to assign the maximum number of fibers within each group. We use multiprocessing to reduce the computing time because this process is the most time-consuming part. When the number of available CPU cores exceeds the number of groups, processing multiple groups simultaneously rather than sequentially reduces the computing time; we use eight cores in this step.
During the optimization, we consider not only the fibers that belong to a group but also nearby fibers that meet certain conditions. However, the use of multiprocessing in this step (i.e. Section~\ref{sec:3.3.1} Process 3-1) can introduce complications, such as different cores assigning different targets to the same fibers or fibers from different groups colliding. Thus, the subsequent step (i.e. Section~\ref{sec:3.3.2} Process 3-2) is necessary to resolve these issues and ensure an optimal solution.

\subsubsection{Process 3-1: Find optimal configuration in each group and its surrounding fibers}\label{sec:3.3.1}
In this process, if the number of possible fiber-target combinations does not exceed the empirical upper limit (i.e. $\sim 10^7$) in a group, we further add neighboring fibers to the group when the assigned target is located in the patrol region of the original group. However, if the number of combinations exceeds this upper limit, we do not add neighboring fibers to avoid the excessive computing time. In the group, we compute all possible fiber-target pairs in this step to find the configuration that maximizes the number of assigned targets without collisions. If multiple valid configurations exist, we select the one that is closest to the fiber-target set assigned in the process 1.

To determine whether a collision occurs between fibers within a group, we define a fiber-target vector $\mathbf{F}^a\in \mathbb{Z}^n$, where $n$ is the number of fibers in the group. Here the $i$-th element $\left\{F^a\right\}_i$ represents either the target ID (i.e. the index of the target) assigned to the $i$-th fiber in the group or zero if no target is assigned. The superscript $a$ denotes the index of possible combinations, where the maximum number of combinations is approximately $\prod_i \left(N_i+1\right)$, with $N_i$ indicating the number of targets assignable to the $i$-th fiber. We also define a vector $\mathbf{C}^b\in \mathbb{Z}^n$, which represents the fiber-target pairs in collision. The superscript $b$ indicates the index of a possible two-fiber collision. For example, if $i$-th and $j$-th fibers are in collision, the vector element $\left\{C^b\right\}_k = 0$ if $k \neq i, \ j$ and $\left\{C^b\right\}_i$, $\left\{C^b\right\}_j$ are the indices of the targets assigned to the $i$-th and $j$-th fibers in collision, respectively. This changes the problem into linear algebra because all the configurations of fibers in collision can be decomposed into sets of two fiber-target pairs in collision. Therefore, if the fibers within an $a$-th configuration do not collide with each other, then they obey the following condition,

\begin{equation}
\label{eq2}
\forall b, \quad \left( (\mathbf{F}^a)^\mathbf{T} -  (\mathbf{C}^b)^\mathbf{T} \right) \cdot \mathbf{C}^b \neq 0 .
\end{equation}
\begin{figure}[t]
    \centering
    \includegraphics[width=0.45\textwidth]{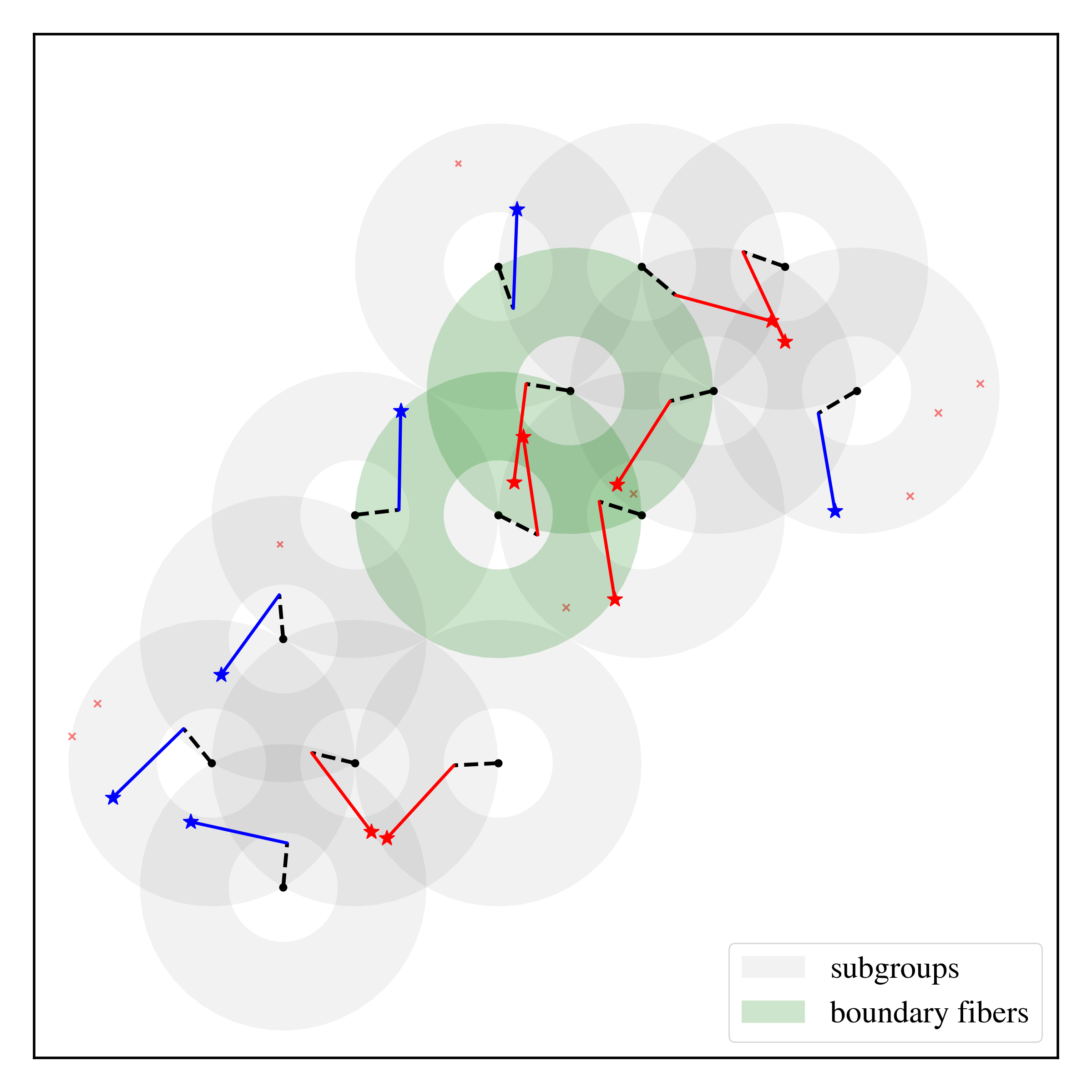}
    \caption{An example of a group splitting into two subgroups and boundary fibers. The black dashed lines and colored solid lines represent the alpha and beta arms. The red fiber positioners are in collision, whereas the blue fiber positioners are not. The dot symbols represent fiber base positions, while the star and cross symbols represent assigned and unassigned targets. The gray shades represent patrol regions of fiber positioners, while the green shades highlight the patrol regions of the boundary fibers.}
    \label{fig: subgroup}
\end{figure}
Because we consider all possible configurations in this process, the computational time increases exponentially with the number of fibers and targets. This issue can be mitigated by implementing an additional strategy that splits the group into two independent subgroups. As shown in Figure~\ref{fig: subgroup}, an ``8"-shaped group (i.e. two groups are connected by a small number of fibers between the groups) can be divided into two subgroups and a boundary (i.e. the connecting fibers); these subgroups are mutually dependent only by the boundary fibers between them. Computing the possible configurations at the boundary, along with the configurations of both groups for each boundary case, significantly reduces the total number of configurations.
This reduction can be approximated as $\left(N+1\right)^l \rightarrow \ \sim 2\times \left(N+1\right)^{l/2}$, where $N$ is the number of assignable targets for each fiber, and $l$ is the number of fibers in the group. We select the boundary fibers in one subgroup that can collide with another fiber in a different subgroup. To do that, we employ the K-means clustering algorithm \citep{Jin2010} to divide the group into two subgroups and to minimize the number of boundary fibers. The K-means algorithm divides a group into subgroups by iteratively assigning each fiber to the subgroup whose centroid is closest, and then updating the subgroup centroids. Minimizing the number of boundary fibers is crucial because the boundary configurations are computed with a ``for loop" in the code, whereas the subgroup configurations are computed with an ``array"; reducing the number of boundary fibers leads to a reduction in computing time. We apply this strategy when both the number of possible combinations and the number of fibers in a group exceed the empirical upper limit (i.e. fiber-target combinations > $10^6$ and the number of fibers > 9). Otherwise, we compute all possible combinations without splitting the group.

\subsubsection{Process 3-2: Find optimal configuration for inconsistent or colliding fibers}\label{sec:3.3.2}

In the previous process, we also consider the surrounding fibers when computing the optimal configuration for each group. However, when multiple groups are computed simultaneously, several issues may arise; different cores may assign different targets to the same fiber, the same target may be assigned to multiple fibers by different cores, or fibers from different groups may collide. The process in this step aims to resolve these issues. 

First, we group the fibers whose assigned targets from the previous process are inconsistent or in collision. The grouping strategy in this process differs from that of the previous one, because neighboring fibers have already been computed in the previous step. Specifically, fibers are grouped if one fiber positioner could collide with another not merely due to their physical separation, but because of the possible fiber assignment configuration; if a fiber positioner can reach the patrol region of another but no fiber-target pairs would result in a collision, they are not grouped. This strategy reduces the group size by splitting them. 
After that, we compute the optimal configuration by repeating the same process (i.e. Process 3-1), but without including neighboring fibers.

\section{Performance}\label{sec : results}

\begin{figure*}[t!]
    \centering
    \includegraphics[width=1\textwidth]{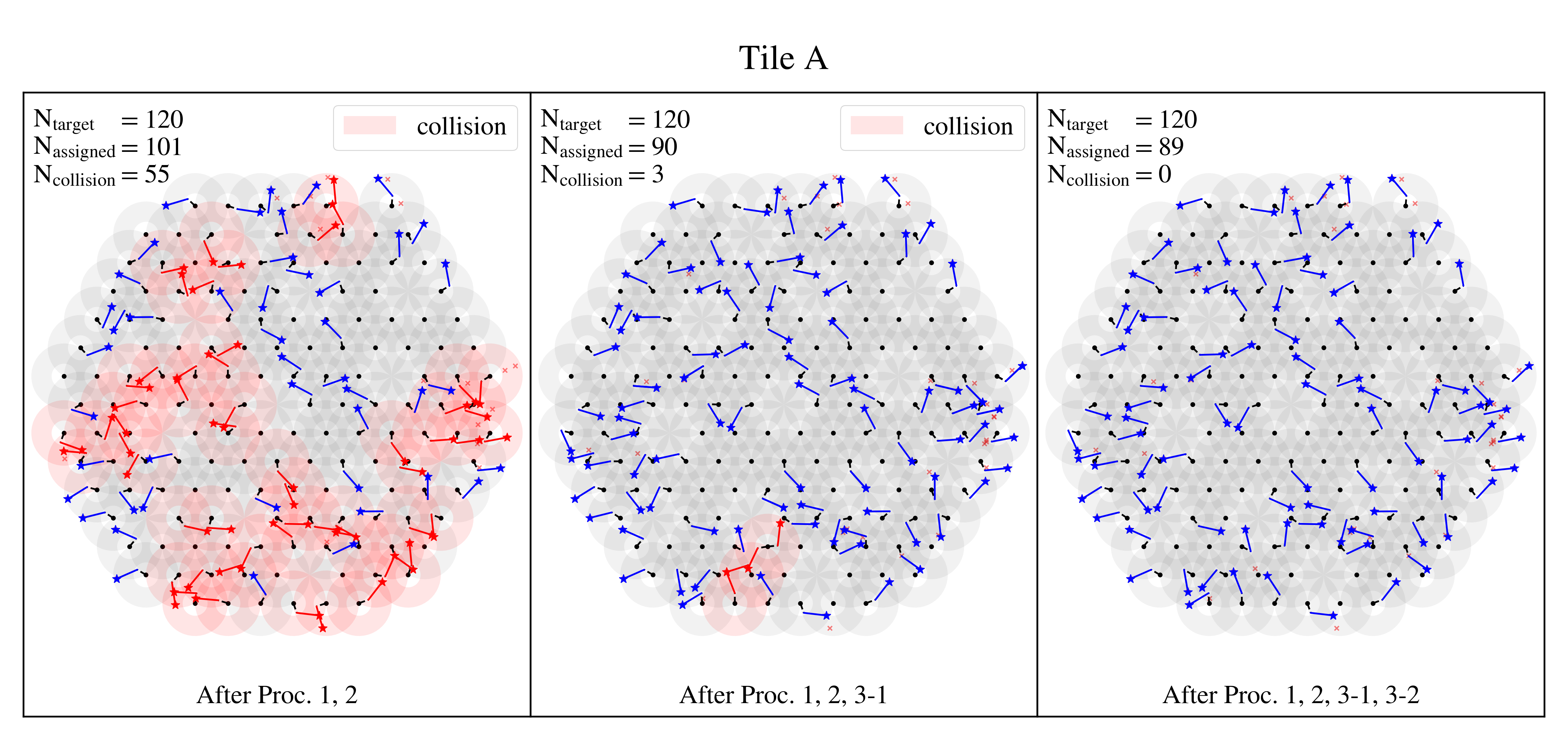}\\
    \centering
    \includegraphics[width=1\textwidth]{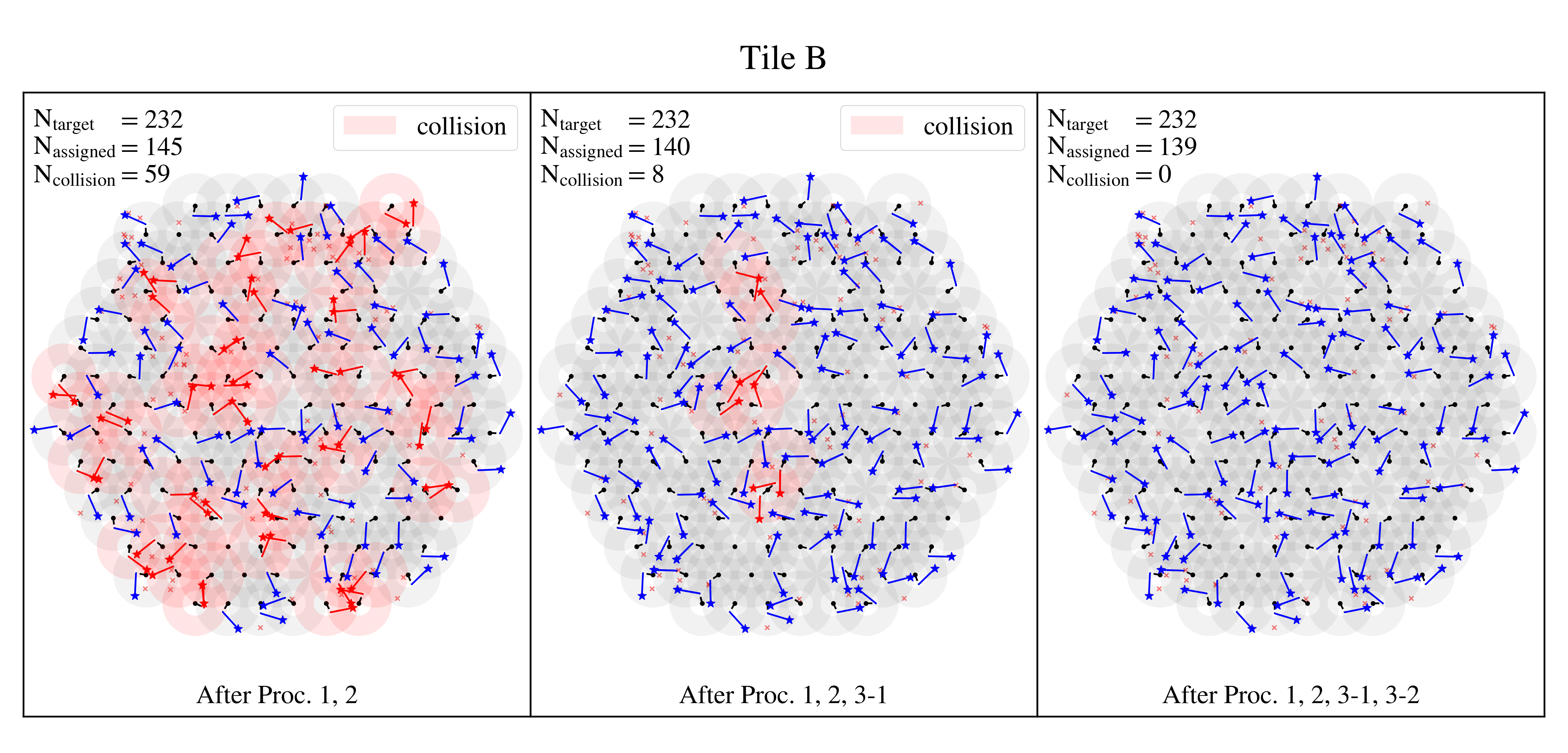}
    \caption{Two examples of fiber assignment results with sample targets from the A-SPEC survey. The red shades describe the patrol regions of collided fibers; all other symbols and shades are as in Figure~\ref{fig: subgroup}. The top and bottom panels show sample assignments with 120 and 232 targets within the fiber patrol regions, respectively. From left to right, the panels show the results after processes 1 $\&$ 2, processes 1, 2 $\&$ 3-1, and after all processes (1, 2, 3-1 $\&$ 3-2).}
    \label{fig: sample}
\end{figure*}

We present two examples of the assignment results in Figure~\ref{fig: sample}; the top and bottom panels indicate different cases. The left panels show the results after processes 1 and 2, where the beta arms in collision are marked as red solid lines. The central panels show the results after process 3-1, and the right panels show the results after all processes. There are fibers that remain in collision after process 3-1 due to inconsistent outcomes from multiprocessing, whereas no fiber positioners remain in collision once all processes are completed.
To demonstrate the performance of our algorithm, we also test another simple algorithm for comparison. Here the simple algorithm assigns targets sequentially from the highest rank while avoiding collisions between fibers.
\begin{figure}[t]
    \centering
    \includegraphics[width=0.5\textwidth]{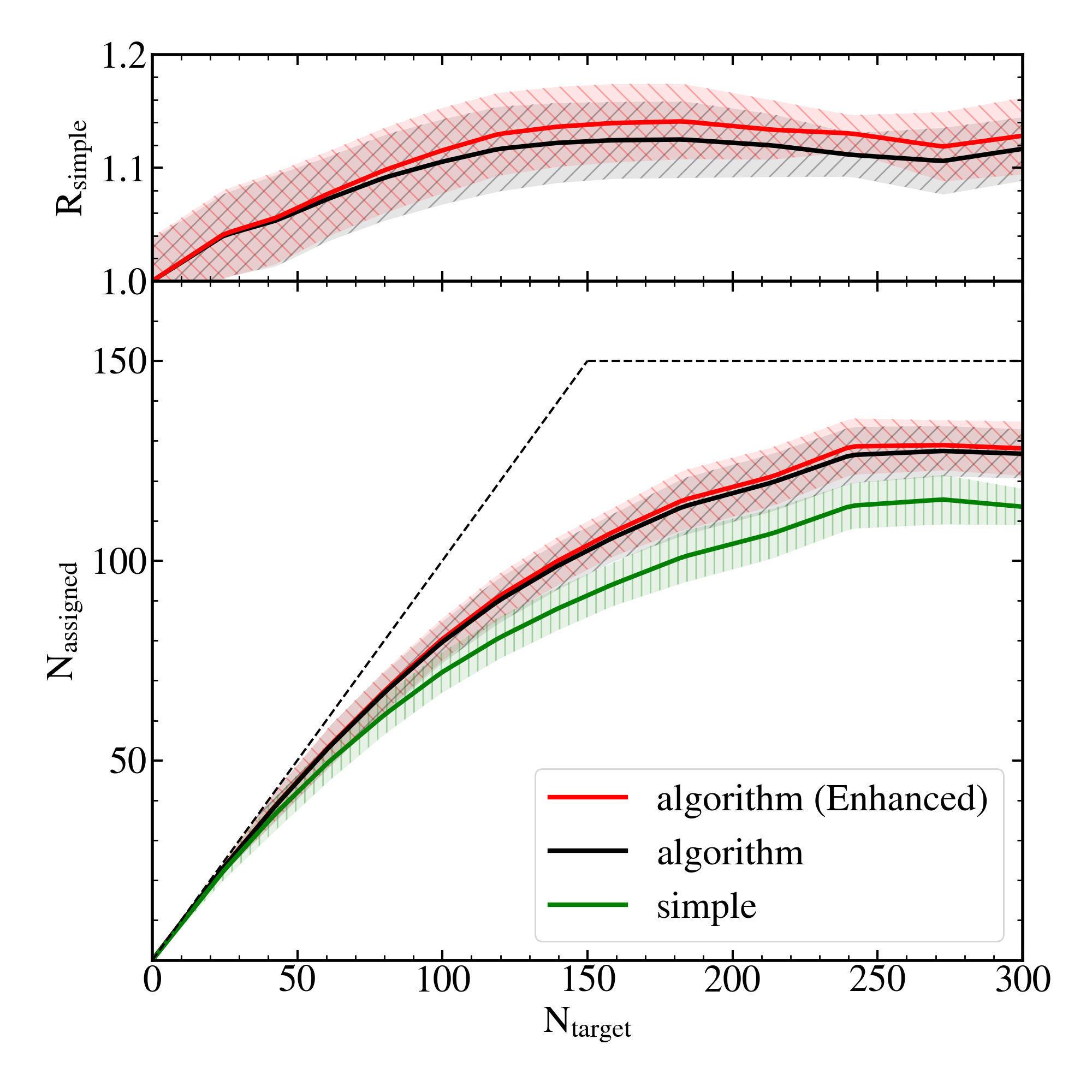}
    \caption{Performance comparison of different algorithms. $\it top$: The ratio of the number of assigned targets using our basic (black) and enhanced (red) algorithms to those assigned with the simple algorithm. $\it bottom$: the numbers of assigned targets as a function of total number of targets within the fiber patrol regions: The green line indicates the result from simple algorithm, whereas black and red solid lines indicate the result from our basic and enhanced algorithms, respectively. The shaded region represents $1\sigma$ standard deviation of the results. The black dashed line indicates the upper bound set by the number of targets and fibers.}
    \label{fig: performance}
\end{figure}

We conduct a test to compare the performance between our and the simple algorithms using 6000 tiles planned for the A-SPEC survey; a tile refers to a single telescope pointing in the MOS survey. In this computation, we use an Intel(R) Xeon(R) Gold 6330 CPU with eight cores. Our algorithm could find, on average, the optimal solution within a second per field and a few minutes even in the most densely clustered regions. The bottom panel of Figure~\ref{fig: performance} shows the numbers of assigned targets from the simple (green) algorithm and our (black) algorithm. The top panel shows how effectively our algorithm improves the target assignment compared to the simple algorithm (i.e. $\rm R_{simple}$ is defined as the number of assigned targets from our algorithm divided by the number of assigned targets from the simple algorithm), plotted as a function of the number of targets. On average, our algorithm assigns 10$\%$ more targets. Also, the ratio reaches its maximum at the point where the number of targets approaches the total number of fibers (i.e. $N_{\rm{fiber}}=150$ in a field).
The average completeness (i.e. the number of assigned targets divided by the number of available targets in the fiber patrol regions) of the A-SPEC sample using our algorithm is $\sim 80\%$, but it drops to $\sim 50 \%$ in a region with the highest target density. The $\sim 80\%$ completeness can be increased up to $\sim 90\%$ when incorporating redshifts already available in the literature; about $40\%$ of the nearby galaxies with $K_S \leq 13.75$ have measured redshifts in the literature. The remaining unfilled fibers in a field will be used for sky and/or secondary target observations. We will discuss the completeness in more detail in a separate paper on the A-SPEC survey plan.

Figure~\ref{fig: performance} also shows the result from a slightly modified version of our algorithm (i.e. enhanced mode, indicated as red solid lines). It is important to minimize the computing time in our algorithm because the survey plan requires many trial assignments with different setups (e.g. different positions of pointing centers) to conduct the most efficient survey. However, once the pointing positions are determined, maximizing the number of assigned targets within a field becomes a priority even at the cost of increased computing time for target assignment. In this mode, the algorithm includes two additional steps, which are explained below; these allow us to assign more targets within a given field even though the computational time is generally increased by a factor of ten compared to the basic mode.

As the first additional step, before executing process 1 in Section~\ref{sec:3.1}, we divide all fiber positioners into groups based on possible collision as in process 3-2 of Section~\ref{sec:3.3.2}. Within these groups, if the possible fiber-target combinations do not exceed the empirical upper limit (i.e. $\sim 10^7$), the algorithm finds the optimal configuration out of all possible cases. These results are completely optimized; therefore, we keep this result and run processes 1 to 3 on the remaining fibers.

During the process 3-1 for the remaining fibers, we consider the surrounding fibers. In the second additional step of the enhanced mode, we adaptively increase the number of these additional fibers until it reaches the upper limit. This adjustment results in larger groups during the process 3-1, which in turn leads to more targets to be assigned. However, it is important to note that the enhanced mode, on average, only assigns one additional target per field compared to the basic mode (with a maximum of four additional targets in our tests) because the basic mode is already highly optimized.

\section{Conclusions}
We have developed a new fiber assignment algorithm designed for a fiber-fed multi-object spectrograph with robotic fiber positioners. The algorithm consists of three steps. First, it assigns the maximum number of targets without considering collisions between fibers. Then, the fibers in collision are grouped. Finally, the algorithm finds the optimal solution for each group and its surrounding fibers.

The performance test using the K-SPEC configuration with potential targets of the A-SPEC survey shows that the overall completeness reaches $\sim 80\%$, which is $\sim 10\%$ higher than the result from the simple algorithm. As more than $40\%$ of the A-SPEC target candidates already have measured redshifts from the literature, the combined completeness can reach $\sim 90\%$ which is the goal of the survey. The average computing time for assignment per field is one second. In summary, our algorithm is computationally efficient and achieves high completeness, making it applicable to other MOS systems with heavily overlapping two-armed fiber positioners.

In practice, the observing fields (i.e. tiles) are selected prior to fiber assignment. After assigning fibers to targets, we compute trajectory plans for all fibers, which describe the motion sequences to the desired positions without collisions \citep[Optimized Step Choice,][]{2025Lim}. In addition, a metrology system measures the fiber-tip positions after these motion sequences are executed and iteratively corrects residual errors to meet the positioning tolerance (Kim D. et al. in prep.). Together, these steps connect the assignment outputs to the instrument hardware and will enable spectroscopic observations in the upcoming A-SPEC survey. This system will be tested during the commissioning observations in February of 2026.


\acknowledgments
We thank the referee for constructive comments.
HSH acknowledges the support of the National Research Foundation of Korea (NRF) grant funded by the Korea government (MSIT), NRF-2021R1A2C1094577, Samsung Electronic Co., Ltd. (Project Number IO220811-01945-01), and Hyunsong Educational \& Cultural Foundation.
This work was partially supported by the Korea Astronomy and Space Science Institute under the R\&D program (Project No. 2025-1-831-00), supervised by the Korea Aerospace Administration.
MH acknowledges financial support under the KASI-KIPAC Fellowship in-kind contribution to Rubin Observatory.
DK is supported by the Global-LAMP Program of the National Research Foundation of Korea (NRF) grant funded by the Ministry of Education (No. RS-2023-00301976).
CBP is supported by the KIAS Individual Grant PG016904 at the Korea Institute for Advanced Study (KIAS) and by the National Research Foundation of Korea (NRF) grant funded by the Korean government (MSIT; RS-2024-00360385).
A-SPEC is an all-sky spectroscopic survey of nearby galaxies conducted with the K-SPEC instrument. A-SPEC is managed by the Korean Spectroscopic Survey Consortium for the Participating Institutions including Korea Astronomy and Space Science Institute (KASI), Korea Institute for Advanced Study (KIAS), and Seoul National University (SNU).




\bibliography{ms.bib}


\end{document}